\documentclass[aps,floatfix,twocolumn,prb,superscriptaddress,reprint,amsmath,amssymb]{revtex4-2}

\usepackage{graphicx}
\usepackage{dcolumn}
\usepackage{bm}

\usepackage[utf8]{inputenc}
\usepackage[T1]{fontenc}
\usepackage{mathptmx}
\usepackage{color}
\usepackage{etoolbox}
\usepackage{braket}
\usepackage{amsmath}
\usepackage{dsfont}
\usepackage{mathtools}
\usepackage{comment}
\usepackage{linop}

\usepackage{array}

\newcommand{\abs}[1]{\ensuremath{\left\vert#1\right\vert}}

\newcommand{\tr}{\text{tr}}
\newcommand{\nh}[1]{#1}

\usepackage{xcolor}
\definecolor{chalmerspurple}{RGB}{103,70,235}
\usepackage{hyperref}
\hypersetup{
  colorlinks,
  citecolor=chalmerspurple,
  linkcolor=chalmerspurple,
  urlcolor=chalmerspurple}

\begin{document}

\title[]{\nh{Superconductivity in a Two-Orbital Hatsugai-Kohmoto Model at Half Filling}}

\author{Nico Hahn}
\affiliation{Department of Physics and Astronomy, Chalmers University of Technology, Gothenburg, Sweden}
\email{nico.hahn@chalmers.se}

\author{R. Matthias Geilhufe}
\affiliation{Department of Physics and Astronomy, Chalmers University of Technology, Gothenburg, Sweden}

\date{\today}

\begin{abstract}
Multiband superconductivity gives rise to a rich landscape of possible pairing states. Here we study superconductivity \nh{emerging from a half-filled normal state in a two-orbital extension} of the Hatsugai-Kohmoto model, an exactly solvable model of correlated electrons with momentum-local interactions, which provides a minimal framework to explore the interplay of strong correlations, orbital structure and pairing symmetry. \nh{For a} system with point-group symmetry $\rm D_{4h}$, we classify the symmetry-allowed superconducting gap structures, taking into account spin, orbital and momentum degrees of freedom. We further compute the critical temperature and the superconducting order parameter for selected pairing channels as functions of interaction and pairing strength within a mean-field treatment. Our results provide a systematic framework for analyzing superconductivity in the orbital Hatsugai-Kohmoto model and extend symmetry-based approaches to correlated multiband settings.
\end{abstract}

\maketitle

\section{Introduction}
\label{secI}

Multiband superconductivity occurs when multiple bands at the Fermi energy contribute to the pairing mechanism. The archetype of this phenomenon is MgB$_2$, which exhibits two distinct superconducting gaps arising from its $\sigma$- and $\pi$-bands \cite{Souma2003}. While MgB$_2$, which features s-wave spin-singlet symmetry, is well described within a modified BCS framework \cite{Suhl1959, Choi2002}, multiband superconductivity in general allows for a richer symmetry landscape than in the single-band case. This enlarged structure is also central to odd-frequency superconductivity \cite{Linder2019}, where multiband degrees of freedom provide additional channels for odd-frequency pairing \cite{BlackSchaffer2013, Triola2020}.

On the other hand, strong correlations play an essential role in many unconventional superconductors. Among those, the iron-based pnictides are probably the most prominent family that also shows multiband effects \cite{Raghu2008, Hirschfeld2011, Chubukov2012}. Further notable systems, see Ref. \onlinecite{Zehetmayer2013} for an overview, include heavy-Fermion superconductors \cite{Rourke2005, Seyfarth2005, Hill2008}, transition-metal dichalcogenides \cite{Majumdar2020, Noat2015} and potentially even metal-organic frameworks, such as the kagome-material Cu-BHT, whose low-energy electronic structure and superconductivity have been discussed both experimentally and theoretically \cite{Zhang2017, Huang2018, Takenaka2021, Ohlrich2025}. Understanding the interplay between multiband effects and strong correlations is therefore of considerable interest.

In recent years, there has been renewed interest in the Hatsugai-Kohmoto (HK) model, an exactly solvable model of correlated electrons \cite{HK1992, HK1998}. Its interaction term is local in momentum space, i.e. of infinite range in real space, which leads to a decoupling of the Hamiltonian into independent momentum sectors and ultimately to exact solvability. Despite its simplicity, the model shows a Mott transition when the interaction strength exceeds a critical value. In Ref. \onlinecite{Huang2022}, this transition was traced back to the breaking of a $\mathbb{Z}_2$ symmetry and it was argued that the HK model is the minimal model to exhibit this symmetry breaking. The model has been explored in various contexts, including superconductivity and competing orders \cite{Phillips2020, Li2022, Zhao2022, Bacsi2025, Corsino2025}, topology \cite{Mai2023A, Mai2023B, Mai2024}, quantum oscillations \cite{Leeb2023, Zhong2024}, and transport and charge response \cite{Ma2025, Guerci2025}. A recent review is given in Ref. \onlinecite{Zhao2025}.

In the present work, we consider the multiband extension of the HK model, also referred to as the orbital HK model. This extension has been studied previously \cite{Bradlyn2023, Mai2024, Mai2026, Tenkila2025}, but not in the context of multiband superconductivity. To introduce superconductivity, we add a pairing term that is treated within a mean-field approximation. This preserves the momentum-local structure so that the resulting mean-field Hamiltonian can be diagonalized in each momentum sector. This procedure has previously been applied to the single-band HK model for s-wave spin-singlet pairing \cite{Phillips2020, Li2022, Zhao2022}. 

We restrict our analysis to two-orbital systems. As a guiding example, we consider nearest- and second-nearest-neighbour hopping between p-orbitals on the square lattice, giving rise to two bands. Up to the orbital interpretation, the non-interacting Hamiltonian is formally equivalent to the minimal two-band model discussed in Ref.~\onlinecite{Raghu2008} for d-orbitals of iron-based superconductors. In comparison to the single-band case, the symmetry classification becomes significantly more extensive due to the inclusion of the orbital degree of freedom alongside spin and momentum. The pairing can thus be a spin-singlet or triplet and an orbital-singlet or triplet under the overall antisymmetry constraint. We perform a classification of pairing symmetries with respect to the point group $\rm D_{4h}$, thereby generalizing the results of Ref. \onlinecite{SigristUeda1991} to the two-orbital setting. The resulting basis functions of the corresponding irreducible representations are the symmetry-allowed candidates for the gap functions. For selected pairing channels, we compute the critical temperature and the superconducting order parameter as functions of interaction and pairing strength.

The outline of this paper is as follows. In Sec.~\ref{secII}, we introduce the HK model and its orbital extension. In Sec.~\ref{secIII}, we classify the symmetry-allowed gap functions for p-orbitals on the square lattice. In Sec.~\ref{secIV}, we discuss the free-energy landscapes and compare the critical temperatures of selected pairing channels. We summarize and conclude in Sec.~\ref{secV}.

\raggedbottom

\section{Framework and Model}
\label{secII}

The HK model, as originally introduced in Ref. \onlinecite{HK1992}, is given by
\begin{equation}
    \op{H}{\text{HK}} = \sum_{\mathbf{k}, \sigma} \left( \varepsilon(\mathbf{k}) - \mu \right) \op{n}{\mathbf{k} \sigma} + U \sum_\mathbf{k} \op{n}{\mathbf{k} \uparrow} \op{n}{\mathbf{k} \downarrow}, 
\end{equation}
where $\op{n}{\mathbf{k} \sigma} = \hc{c}{\mathbf{k} \sigma} \op{c}{\mathbf{k} \sigma}$ is the particle number operator for momentum $\mathbf{k}$ and spin $\sigma$, and $\op{c}{\mathbf{k} \sigma}$, $\hc{c}{\mathbf{k} \sigma}$ denote the corresponding electronic annihilation and creation operators. The model describes a single spin-degenerate band $\varepsilon(\mathbf{k})$ and is therefore sometimes referred to as the band HK model, in distinction to its orbital generalization introduced below.

Its key feature is the momentum-local interaction term, which allows for a decomposition of the Hamiltonian into decoupled $\mathbf{k}$-sectors, rendering the model exactly solvable. In real space, the momentum-local interaction corresponds to an all-to-all interaction of infinite range \cite{Skolimowski2024}, reminiscent of other exactly solvable models such as the SYK model for correlated Fermions \cite{Rosenhaus2019} and the Sherrington-Kirkpatrick model for spin glasses \cite{Panchenko2012}.

Despite its simplicity, the model shows a Mott transition at a critical interaction strength $U_C > 0$ determined by the bandwidth. In Ref. \onlinecite{Huang2022}, this transition was traced back to the breaking of a $\mathbb{Z}_2$-symmetry and the HK model was argued to constitute the minimal model capable of such a symmetry breaking. Motivated by this, we adopt the HK model as the minimal framework for studying strongly correlated electron physics. However, the band HK model features a thermodynamic degeneracy in the ground state and a diverging magnetic susceptibility. As shown in Ref.~\onlinecite{Bradlyn2023}, these unphysical features need not persist in orbital extensions of the model and can be removed in suitable multiorbital generalizations.

The general form of the orbital HK model is
\begin{equation} \label{2OrbitalHK}
\begin{split}
    \op{H}{\text{OHK}} = &\sum_{\mathbf{k}, \alpha, \beta, \sigma} \left( H_{\alpha \beta}(\mathbf{k}) - \mu \delta_{\alpha \beta} \right) \hc{c}{\mathbf{k} \alpha \sigma} \op{c}{\mathbf{k} \beta \sigma} 
    \\
    &+ \sum_{\mathbf{k}, \alpha, \beta}  U_{\alpha \beta}(\mathbf{k}) \op{n}{\mathbf{k} \alpha \uparrow} \op{n}{\mathbf{k} \beta \downarrow},
\end{split}
\end{equation}
where an orbital index $\alpha$ is added to the electronic operators. We assume the Bloch Hamiltonian $H(\mathbf{k})$ to be spin-degenerate, i.e. spin-orbit coupling is absent. Furthermore, we consider an interaction that is isotropic in momentum space and diagonal and orbital-independent in the orbital basis, $U_{\alpha \beta}(\mathbf{k}) = U \delta_{\alpha \beta}$. This minimal choice excludes more general momentum-local multiorbital interactions, such as interorbital density terms, Hund coupling and 
pair hopping. Relaxing the isotropy condition has been shown to induce Fermi arcs and pseudogaps in the band HK model \cite{Yang2021, Yang2023}.

For the following analysis, we specify a two-band model for the Bloch Hamiltonian $H(\mathbf{k})$. We consider the square lattice with two p-orbitals per site and include nearest- and second-nearest-neighbour hopping. This leads to the Bloch \mbox{Hamiltonian}
\begin{equation}    \label{2BlochHamiltonian}
\begin{split}
    H(\mathbf{k}) =& \left[ 4s \cos k_x \cos k_y + \left( t_\sigma^{(1)} + t_\pi^{(1)} \right) \left( \cos k_x + \cos k_y \right) \right] \tau_0
    \\
    &- 4r \sin k_x \sin k_y \tau_x 
    + \left( t_\sigma^{(1)} - t_\pi^{(1)} \right) \left( \cos k_x - \cos k_y \right) \tau_z,
\end{split}
\end{equation}
where $\tau_x$, $\tau_y$, and $\tau_z$ are the Pauli matrices acting in orbital space and $\tau_0$ is the identity matrix.

The parameters are determined by the overlap integrals \cite{SlaterKoster1954} $t_\sigma^{(1)} = (\rm{pp}\sigma)_1, t_\pi^{(1)} = (\rm{pp}\pi)_1, s = 
((\rm{pp}\sigma)_2 + (\rm{pp}\pi)_2)/2, r = ((\rm{pp}\sigma)_2 - (\rm{pp}\pi)_2)/2$. The same tight-binding model has been applied in Ref. \onlinecite{Raghu2008} to describe nearest-neighbour hopping between the d-orbitals of Fe atoms in iron-based superconductors. \nh{This correspondence is limited to the structure of the non-interacting two-orbital Hamiltonian and does not extend to material-specific parameters or interactions.} The present p-orbital model is discussed in more detail in Appendix~\ref{A1}. In this specific two-band model, the aforementioned thermodynamic ground-state degeneracy of the band HK model is retained. We use this
model not as a multiorbital mechanism for lifting the normal-state degeneracy, but as a minimal two-orbital setting for classifying superconducting pairing channels.

On top of this normal-state model, we incorporate superconductivity by adding a pairing term
\begin{equation}    \label{2FullHamiltonian}
    \op{H} = \op{H}{\text{OHK}} - \frac{g}{N} \hc{A} \op{A},
\end{equation}
where $N$ denotes the number of unit cells or, equivalently, the number of $\mathbf{k}$-points in the Brillouin zone. This form follows Refs.~\onlinecite{Phillips2020, Li2022, Zhao2022}. We restrict ourselves to pairing states with zero center-of-mass momentum. The pairing operator can be written in the general form
\begin{equation}    \label{2PairingOperator}
    \op{A} = \sum_\mathbf{k} \op{\psi}{-\mathbf{k}}{T} W(\mathbf{k}) \op{\psi}{\mathbf{k}},
\end{equation}
where the $2n$-vector
\begin{equation}
    \op{\psi}{\mathbf{k}} = \left( \op{c}{\mathbf{k} \alpha_1 \uparrow}, \ldots, \op{c}{\mathbf{k} \alpha_n \uparrow}, \op{c}{\mathbf{k} \alpha_1 \downarrow}, \ldots, \op{c}{\mathbf{k} \alpha_n \downarrow}\right)
\end{equation}
contains the electronic annihilation operators in a suitable $n$-dimensional basis. In the present work, we choose the orbital basis. The gap function $W(\mathbf{k}) \in \mathbb{C}^{2n \times 2n}$ encodes the symmetry of the superconducting state. For $\op{A}$ to be non-vanishing, $W(\mathbf{k})$ must satisfy $W(\mathbf{k}) = -W^T(-\mathbf{k})$. Due to the square geometry, the model \eqref{2BlochHamiltonian} has the symmetry of the point group $\rm D_{4h}$. This constrains the possible gap functions $W(\mathbf{k})$ and provides the basis for the symmetry classification in the following section.

\section{Symmetry Classification}
\label{secIII}

\begin{table}[t!]
    \begin{tabular}{|>{\centering\arraybackslash}p{1.4cm}|>{\centering\arraybackslash}p{1.4cm}|>{\centering\arraybackslash}p{1.4cm}|>{\centering\arraybackslash}p{3.8cm}|}
    \hline
         Spin & Parity & Orbital & $W(\mathbf{k})$
         \\
    \hline
    \hline
         Singlet & Odd & Singlet & $\varphi(\mathbf{k}) i \sigma_y \otimes i \tau_y$
         \\
         Singlet & Even & Triplet & $i \sigma_y \otimes \left[\boldsymbol{\ell}(\mathbf{k}) \cdot \boldsymbol{\tau} \right] i \tau_y$
         \\
         Triplet & Even & Singlet & $\left[\mathbf{d} (\mathbf{k}) \cdot \boldsymbol{\sigma} \right]i \sigma_y \otimes i \tau_y$
         \\
         Triplet & Odd & Triplet & $\left[\mathbf{d} (\mathbf{k}) \cdot \boldsymbol{\sigma} \right]i \sigma_y \otimes \left[\boldsymbol{\ell} (\mathbf{k}) \cdot \boldsymbol{\tau} \right]i \tau_y$
         \\
    \hline
    \end{tabular}
    \caption{Overview of spin and orbital pairing sectors, their parity, and the corresponding structure of the gap function.}
    \label{Tab1}
\end{table}

In the single-band case ($n = 1$), the gap function $W(\mathbf{k})$ acts only in spin space and is therefore a $2\times 2$-matrix,
\begin{equation}
    W(\mathbf{k}) = i \varphi(\mathbf{k}) \sigma_y
    \quad  \text{or}   \quad
    W(\mathbf{k}) = i \left( \mathbf{d}(\mathbf{k}) \cdot \boldsymbol{\sigma} \right)\sigma_y
\end{equation}
for spin-singlet and spin-triplet pairing, respectively. Antisymmetry requires $\varphi(\mathbf{k})$ to be even and $\mathbf{d}(\mathbf{k})$ to be odd under $\mathbf{k} \to -\mathbf{k}$.

In the two-band case ($n = 2$), the orbital degrees of freedom are included in the matrix $W(\mathbf{k})$. This allows for both orbital-singlet and orbital-triplet states. \nh{In analogy to the $\mathbf{d}$-vector in the spin-triplet sector, we parametrize the orbital-triplet sector by an orbital pseudospin vector $\boldsymbol{\ell}(\mathbf{k})$ through the symmetric matrix $\left[\boldsymbol{\ell}(\mathbf{k})\cdot\boldsymbol{\tau}\right]i\tau_y$, while the orbital-singlet sector is proportional to $i\tau_y$.} Denoting $S$ as behaviour under spin exchange ($S = -1$ for the antisymmetric singlet state and $S = +1$ for the symmetric triplet state), $O$ as behaviour under orbital exchange and $P$ as parity, we obtain the rule
\begin{equation}
    S\, P\, O = -1.
\end{equation}
The resulting pairing sectors are summarized in Tab.~\ref{Tab1}.

Having established the four spin-orbital parity sectors allowed by Fermi antisymmetry, we now classify the corresponding gap structures for the point group $\rm D_{4h}$. Under the action of $h \in \rm D_{4h}$, the gap function $W(\mathbf{k})$ transforms as
\begin{equation}
    W(\mathbf{k}) \to \left[S(h) \otimes O(h)\right] W(R_-^{-1}(h) \mathbf{k}) \left[S(h) \otimes O(h) \right]^T,
\end{equation}
where $S(h) \in \rm SU(2)$ is the spinor representation acting on spin space, $O(h) \in \rm E_u \subset U(2)$ is the orbital representation and $R_\pm(h) \in \rm O(3)$ are the three-dimensional vector representations. Here $O(h)$ belongs to the two-dimensional representation $\rm E_u$ of $\rm D_{4h}$, which is the orbital representation for the p-orbitals in our model. We distinguish the three-dimensional representations $R_\pm(h) \in \rm O(3)$ based on their behaviour under inversion: polar vectors, such as the momentum $\mathbf{k}$, transform via $R_-(h)$, while axial vectors, such as the $\mathbf{d}$-vector in the spin-triplet case, transform via $R_+(h)$.

\begin{table}[!t]
    \begin{tabular}{|>{\centering\arraybackslash}p{2.4cm}|>{\centering\arraybackslash}p{2.8cm}|>{\centering\arraybackslash}p{2.8cm}|}
    \hline
         Representation & $W(\mathbf{k})$ & $\varphi(\mathbf{k})$
         \\
         \hline
         \hline
         $\rm A_{1u}$ 
         & 
         $z\, \sigma_y \otimes \tau_y$ & 
         $z$
         \\
         $\rm A_{2u}$ & & 
         \\
         $\rm B_{1u}$ & & 
         \\
         $\rm B_{2u}$ & & 
         \\
         $\rm E_u$ 
         & $x\, \sigma_y \otimes \tau_y$, 
         $y\, \sigma_y \otimes \tau_y$ & 
         $x$, 
         $y$
         \\
    \hline
    \end{tabular}
    \caption{Basis gap functions for the irreducible representations of the point group $\rm D_{4h}$ in the spin-singlet and orbital-singlet sector.}
    \label{Tab2}
\end{table}

For the spin-singlet sector, the action of $\rm SU(2)$ is redundant, since $U^T \sigma_y U = \sigma_y$ for $U \in \rm SU(2)$. In other words, the special unitary group is a subset of the complex symplectic group, $\rm SU(2) \subset Sp(2, \mathbb{C})$. In contrast, for the spin-triplet sector, the spin rotation corresponds to a rotation of the $\mathbf{d}$-vector, $\mathbf{d}(\mathbf{k}) \to R_+(h)\, \mathbf{d}(R_-^{-1}(h) \mathbf{k})$, reflecting the homomorphism between $\rm SU(2)$ and $\rm SO(3)$.

\begin{table}[!t]
\centering

\begin{minipage}{\columnwidth}
    \begin{tabular}{|>{\centering\arraybackslash}p{2.4cm}|>{\centering\arraybackslash}p{2.8cm}|>{\centering\arraybackslash}p{2.8cm}|}
    \hline
        Representation & $W(\mathbf{k})$ & $\boldsymbol{\ell} (\mathbf{k})$
        \\
        \hline
        \hline
        $\rm A_{1g}$ 
        & 
        $\sigma_y \otimes \tau_0$,
        $\left( x^2 + y^2 \right) \sigma_y \otimes \tau_0$,
        $x\, y\, \sigma_y \otimes \tau_x$,
        $\left( x^2 - y^2 \right) \sigma_y \otimes \tau_z$,
        $z^2 \sigma_y \otimes \tau_0$
        & 
        $(0,1,0)$,
        $(0,x^2+y^2,0)$,
        $(0,0,x\, y)$,
        $(x^2-y^2,0,0)$,
        $(0,z^2,0)$
        \\
        $\rm A_{2g}$ 
        & 
        $x\, y\, \sigma_y \otimes \tau_z$, 
        $(x^2-y^2) \sigma_y \otimes \tau_x$
        & 
        $(x\, y,0,0)$, 
        $(0,0,x^2-y^2)$
        \\
        $\rm B_{1g}$ 
        & 
        $\sigma_y \otimes \tau_z$,
        $\left( x^2+y^2 \right) \sigma_y \otimes \tau_z$,
        $\left( x^2-y^2 \right) \sigma_y \otimes \tau_0$,      
        $z^2 \sigma_y \otimes \tau_z$
        & 
        $(1,0,0)$,
        $(x^2+y^2,0,0)$,
        $(0,x^2-y^2,0)$,
        $(z^2, 0, 0)$
        \\
        $\rm B_{2g}$
        & 
        $\sigma_y \otimes \tau_x$, 
        $\left(x^2 + y^2 \right) \sigma_y \otimes \tau_x$,
        $x\, y\, \sigma_y \otimes \tau_0$, 
        $z^2 \sigma_y \otimes \tau_x$   & 
        $(0,0,1)$,
        $(0,0,x^2+y^2)$,
        $(0,x\, y,0)$,
        $(0,0,z^2)$        
        \\
        $\rm E_{g}$ 
        & 
        $x\, z\, \sigma_y \otimes \tau_z$, 
        $x\, z\, \sigma_y \otimes \tau_0$, 
        $x\, z\, \sigma_y \otimes \tau_x$, 
        $y\, z\, \sigma_y \otimes \tau_z$, 
        $y\, z\, \sigma_y \otimes \tau_0$, 
        $y\, z\, \sigma_y \otimes \tau_x$
        & 
        $(x\, z,0,0)$, 
        $(0,x\, z,0)$, 
        $(0,0,x\, z)$,
        $(y\, z,0,0)$, 
        $(0,y\, z,0)$, 
        $(0,0,y\, z)$
        \\
    \hline
    \end{tabular}
    \caption{Basis gap functions for the irreducible representations of the point group $\rm D_{4h}$ in the spin-singlet and orbital-triplet sector.}
    \label{Tab3}
    \end{minipage}

    \vspace{0.5cm}
    
    \begin{minipage}{\columnwidth}
    \begin{tabular}{|>{\centering\arraybackslash}p{2.4cm}|>{\centering\arraybackslash}p{2.8cm}|>{\centering\arraybackslash}p{2.8cm}|}
    \hline
        Representation & $W (\mathbf{k})$ & $\mathbf{d} (\mathbf{k})$
        \\
        \hline
        \hline
        $\rm A_{1g}$ 
        & 
        $\sigma_x \otimes \tau_y$,
        $\left( x^2 + y^2 \right) \sigma_x \otimes \tau_y$,
        $z \left( y\, \sigma_0 + i x\, \sigma_z \right) \otimes \tau_y$,        
        $z^2 \sigma_x \otimes \tau_y$ 
        & 
        $(0, 0, 1)$,
        $(0,0,x^2+y^2)$,
        $(x\, z, y\, z,0)$,
        $(0,0,z^2)$
        \\
        $\rm A_{2g}$ 
        & 
        $z \left( x\, \sigma_0 - i y\, \sigma_z \right) \otimes \tau_y$ 
        & 
        $(-y\, z, x\, z, 0)$
        \\
        $\rm B_{1g}$ 
        & 
        $z \left( x\, \sigma_z + i y\, \sigma_0 \right) \otimes \tau_y$, 
        $(x^2-y^2) \sigma_x \otimes \tau_y$ 
        & 
        $(-x\, z, y\, z, 0)$, 
        $(0,0,x^2-y^2)$
        \\
        $\rm B_{2g}$ 
        & 
        $z \left( x\, \sigma_0 + i y\, \sigma_z \right) \otimes \tau_y$, 
        $x\, y\, \sigma_x \otimes \tau_y $
        & 
        $(y\, z, x\, z, 0)$, 
        $(0,0,x\, y)$
        \\
        $\rm E_{g}$ 
        & 
        $\sigma_0 \otimes \tau_y$,
        $\sigma_z \otimes \tau_y$,
        $(x^2+y^2) \sigma_0 \otimes \tau_y$,
        $(x^2+y^2) \sigma_z \otimes \tau_y$,
        $z^2 \sigma_0 \otimes \tau_y$,
        $z^2 \sigma_z \otimes \tau_y$,
        $(x^2-y^2) \sigma_0 \otimes \tau_y$,
        $(x^2-y^2) \sigma_z \otimes \tau_y$,
        $x\, y \sigma_0 \otimes \tau_y$,
        $x\, y \sigma_z \otimes \tau_y$,
        $y\, z \sigma_x \otimes \tau_y$, 
        $x\, z \sigma_x \otimes \tau_y$
        &
        $(0,1,0)$,
        $(1,0,0)$,
        $(0,x^2+y^2,0)$,
        $(x^2+y^2,0,0)$,
        $(0,z^2,0)$,
        $(z^2,0,0)$,
        $(0,x^2-y^2,0)$,
        $(x^2-y^2,0,0)$,
        $(0,x\, y,0)$,
        $(x\, y,0,0)$,
        $(0,0,y\, z)$,
        $(0,0,x\, z)$,
        \\
    \hline
    \end{tabular}
    \caption{Basis gap functions for the irreducible representations of the point group $\rm D_{4h}$ in the spin-triplet and orbital-singlet sector.}
    \label{Tab4}
    \end{minipage}
\end{table}

\begin{table*}[!t]
    \begin{tabular}{|>{\centering\arraybackslash}p{2.4cm}|>{\centering\arraybackslash}p{6cm}|>{\centering\arraybackslash}p{4cm}|>{\centering\arraybackslash}p{4cm}|}
    \hline
        Representation & $W(\mathbf{k})$ & $\mathbf{d} (\mathbf{k})$ & $\boldsymbol{\ell}(\mathbf{k})$
        \\
    \hline
    \hline
        $\rm A_{1u}$ 
        & 
        $\left( x\, \sigma_z - i\, y\, \sigma_0 \right) \otimes \tau_0$,
        $z\, \sigma_x \otimes \tau_0$,
        $\left( x\, \sigma_0 + i\, y\, \sigma_z\right) \otimes \tau_x $,
        $\left( x\, \sigma_z + i\, y\, \sigma_0 \right) \otimes \tau_z$
        & 
        $(x,y,0)$, 
        $(0,0,z)$, 
        $(y,x,0)$, 
        $(x,-y,0)$
        & 
        $(0,1,0)$, 
        $(0,1,0)$, 
        $(0,0,1)$, 
        $(1,0,0)$
        \\
        $\rm A_{2u}$ 
        & 
        $\left( x\, \sigma_0 - i\, y\, \sigma_z \right) \otimes \tau_0$,
        $\left( x\, \sigma_z + i\, y\, \sigma_0 \right) \otimes \tau_x$,
        $\left( x\, \sigma_0 + i\, y\, \sigma_z \right) \otimes \tau_z$
        & 
        $(-y,x,0)$, 
        $(x,-y,0)$, 
        $(y,x,0)$
        & 
        $(0,1,0)$, 
        $(0,0,1)$, 
        $(1,0,0)$
        \\
        $\rm B_{1u}$
        & 
        $\left( x\, \sigma_z +i\, y\, \sigma_0 \right) \otimes \tau_0$,
        $\left( x\, \sigma_0 -i\, y\, \sigma_z \right) \otimes \tau_x$,
        $\left( x\, \sigma_z -i\, y\, \sigma_0 \right) \otimes \tau_z$,
        $z\, \sigma_x \otimes \tau_z$ 
        & 
        $(x,-y,0)$, 
        $(-y,x,0)$, 
        $(x,y,0)$, 
        $(0,0,z)$
        & 
        $(0,1,0)$, 
        $(0,0,1)$, 
        $(1,0,0)$, 
        $(1,0,0)$
        \\
        $\rm B_{2u}$ 
        & 
        $\left( x\, \sigma_0 +i\, y\, \sigma_z \right) \otimes \tau_0$,
        $\left( x\, \sigma_z -i\, y\, \sigma_0 \right) \otimes \tau_x$,
        $z\, \sigma_x \otimes \tau_x$,
        $\left( x\, \sigma_0 -i\, y\, \sigma_z \right) \otimes \tau_z$
        & 
        $(y,x,0)$, 
        $(x,y,0)$, 
        $(0,0,z)$, 
        $(-y,x,0)$
        & 
        $(0,1,0)$, 
        $(0,0,1)$, 
        $(0,0,1)$, 
        $(1,0,0)$
        \\
        $\rm E_u$ 
        &
        $x\, \sigma_x \otimes \tau_0$,
        $x\, \sigma_x \otimes \tau_x$,
        $x\, \sigma_x \otimes \tau_z$,
        $y\, \sigma_x \otimes \tau_0$,
        $y\, \sigma_x \otimes \tau_x$,
        $y\, \sigma_x \otimes \tau_z$
        &
        $(0,0,x)$, 
        $(0,0,x)$, 
        $(0,0,x)$, 
        $(0,0,y)$, 
        $(0,0,y)$, 
        $(0,0,y)$
        &
        $(0,1,0)$, 
        $(0,0,1)$, 
        $(1,0,0)$, 
        $(0,1,0)$, 
        $(0,0,1)$, 
        $(1,0,0)$
        \\
        &
        $z\, \sigma_0 \otimes \tau_0$,
        $z\, \sigma_0 \otimes \tau_x$,
        $z\, \sigma_0 \otimes \tau_z$,
        $z\, \sigma_z \otimes \tau_0$,
        $z\, \sigma_z \otimes \tau_x$,
        $z\, \sigma_z \otimes \tau_z$
        & 
        $(0,z,0)$, 
        $(0,z,0)$, 
        $(0,z,0)$, 
        $(z,0,0)$, 
        $(z,0,0)$, 
        $(z,0,0)$
        & 
        $(0,1,0)$, 
        $(0,0,1)$, 
        $(1,0,0)$, 
        $(0,1,0)$, 
        $(0,0,1)$, 
        $(1,0,0)$
        \\
    \hline
    \end{tabular}
    \caption{Basis gap functions for the irreducible representations of the point group $\rm D_{4h}$ in the spin-triplet and orbital-triplet sector.}
    \label{Tab5}
\end{table*}

This equivalence does not carry over to the orbital sector. For the orbital-singlet, one finds $U^T \tau_y U = \det U\ \tau_y$ for $U \in \rm E_u \subset \rm U(2)$. The additional phase depends on the group element and therefore cannot be absorbed into a global $\rm U(1)$ phase when projecting onto irreducible representations using the character projection method \cite{gtpack2}. In other words, the matrices of the representation $\rm E_u$ are not a subset of the complex symplectic group $\rm Sp(2, \mathbb{C})$, but of the projective symplectic group $\rm E_u \subset PSp(2, \mathbb{C})$. In the orbital-triplet sector, the orbital rotation can, in analogy to the rotation in spin space, be performed via a rotation of the $\boldsymbol{\ell}$-vector only when the phase factors are taken into account.

The group $\rm D_{4h}$ has ten irreducible representations, five of which are even and five of which are odd under inversion. We compute the basis functions for the irreducible representations of $\rm D_{4h}$ using the Wolfram Mathematica group theory package GTPack \cite{gtpack1, gtpack2}. We distinguish between the four cases given in Tab.~\ref{Tab1}. The results can be found in Tab.~\ref{Tab2} for the spin-singlet and orbital-singlet sector, in Tab.~\ref{Tab3} for the spin-singlet and orbital-triplet sector, in Tab.~\ref{Tab4} for the spin-triplet and orbital-singlet sector and in Tab.~\ref{Tab5} for the spin-triplet and orbital-triplet sector. These tables list the basis functions allowed by symmetry, expressed as polynomials in the Cartesian coordinates $(x, y, z)$. Overall phase factors are omitted in Tabs.~\ref{Tab2}-~\ref{Tab5}. For a crystal lattice, these functions are replaced by the corresponding lowest-order lattice harmonics, obtained by the substitutions $x \to \sin\, k_x$, $x^2 \to \cos\, k_x$ etc. These substitutions ensure that the functions respect the periodicity of the Brillouin zone while retaining the transformation properties under the point group $\rm D_{4h}$. From these tables we select the gap functions used in the numerical calculations below. For the two-dimensional system at hand, we exclude gap functions with any form of $z$-dependence.

As a consistency check, the internal matrix structure decomposes as
\begin{equation}
    \left( 1 \oplus 3 \right)_\text{spin} \otimes \left( 1 \oplus 3 \right)_\text{orbital},
\end{equation}
corresponding to the four sectors with dimensions $1$, $3$, $3$, and $9$, respectively. The momentum dependence is then supplied by basis functions transforming according to irreducible representations of $\rm D_{4h}$, subject to the constraint $S P O = -1$.

The normal-state Hamiltonian introduced in Sec.~\ref{secII} has a global SU(2) symmetry in spin space. Consequently, gap functions in the spin-triplet sector that are related by a rotation of the $\mathbf{d}$-vector are degenerate, even when they belong to different irreducible representations of $\rm D_{4h}$. This degeneracy is lifted once spin rotation symmetry is broken, for example by spin-orbit coupling or spin-dependent interactions. The point group symmetry enforces degeneracies only within the multidimensional irreducible representations $\rm E_g$ and $\rm E_u$. 

\nh{For the numerical analysis below, we focus on the following gap functions}
\begin{equation}    \label{3SelectedPairingChannels}
\begin{split}
    W_{\rm A_{1g}}^\text{(ST)}(\mathbf{k}) &= \frac{1}{2} \sigma_y \otimes \tau_0,
    \\
    W_{\rm E_u}^\text{(SS)}(\mathbf{k}) &= \frac{1}{\sqrt{2}} \sin k_x\, \sigma_y \otimes \tau_y,
    \\
    \nh{W_{\rm B_{1g}}^\text{(ST)}(\mathbf{k})} &\nh{= \frac{1}{2} \left( \cos k_x - \cos k_y \right) \sigma_y \otimes \tau_0,}
    \\
    \nh{W_{\rm B_{2g}}^\text{(TS)}(\mathbf{k})} &\nh{= \sin k_x\, \sin k_y \sigma_x \otimes \tau_y,}
\end{split}
\end{equation}
where the superscripts SS, ST, and TS indicate the spin-singlet/orbital-singlet, spin-singlet/orbital-triplet and spin-triplet/orbital-singlet, respectively. We also denote these channels as $\rm A_{1g}^{(ST)}$, $\rm E_u^{(SS)}$, \nh{$\rm B_{1g}^{(ST)}$ and  $\rm B_{2g}^{(TS)}$}. \nh{These four channels sample different spin-orbital sectors, parity classes, and momentum dependences. In particular, $\rm A_{1g}^{(ST)}$ is the simplest fully symmetric reference channel and $\rm E_u^{(SS)}$ provides an odd-parity spin- and orbital-singlet example. The two-dimensional representation $\rm E_u$ allows for nematic or chiral two-component order parameters. Here, we restrict the numerical analysis to the single $\mathrm{p}_x$-like component. The channel $\rm B_{1g}^{(ST)}$ is a spin-singlet/orbital-triplet channel with an even-parity nodal, $\mathrm{d}_{x^2-y^2}$-like momentum dependence, and $\rm B_{2g}^{(TS)}$ represents a spin-triplet/orbital-singlet channel with an even-parity nodal, $\mathrm{d}_{x y}$-like momentum dependence.
}

The functions are normalized with the Frobenius norm
\begin{equation}
    \frac{1}{(2\pi)^2} \int_\text{BZ} d^2k\, \tr\, W^\dag(\mathbf{k}) W(\mathbf{k}) = 1.
\end{equation}
In weak-coupling BCS theory, pairing is typically restricted to electronic states within a narrow energy window around the Fermi surface. Here, we consider the strong coupling regime, where all $\mathbf{k}$-points contribute equally.

\nh{
The gap functions can also be characterized in the band basis of the normal-state Bloch Hamiltonian \eqref{2BlochHamiltonian}. It has the form
\begin{equation}
    H(\mathbf{k}) = h_0(\mathbf{k}) \tau_0 + h_x(\mathbf{k}) \tau_x + h_z(\mathbf{k}) \tau_z
\end{equation}
and can therefore be diagonalized by a real orbital rotation
$U(\mathbf{k})$. Since $H(-\mathbf{k}) = H(\mathbf{k})$, it may be chosen such that $U(-\mathbf{k}) = U(\mathbf{k})$. The orbital parts of the selected gap functions \eqref{3SelectedPairingChannels} transform as
\begin{equation}
    U^T(-\mathbf{k}) \tau_0 U(\mathbf{k}) = \tau_0,
    \qquad
    U^T(-\mathbf{k}) \tau_y U(\mathbf{k}) = \det U(\mathbf{k})\, \tau_y,
\end{equation}
where we used $U(-\mathbf{k}) = U(\mathbf{k})$. Thus, $\tau_0$ remains diagonal in the band basis, while $\tau_y$ remains purely off-diagonal up to an irrelevant sign. The former therefore describes intraband pairing and the latter interband pairing. Consequently, the $\rm A_{1g}^{(\mathrm{ST})}$ and $\rm B_{1g}^{(\mathrm{ST})}$ channels are purely intraband, while the $\rm E_u^{(\mathrm{SS})}$ and $\rm B_{2g}^{(\mathrm{TS})}$ channels are purely interband.
}

\section{Mean-Field Theory and Numerical Analysis}
\label{secIV}

A mean-field decoupling of the pairing term preserves the momentum-sector decomposition of the HK model, such that the resulting local Hamiltonians can be diagonalized exactly for each sector. This procedure was previously applied to the band HK model with s-wave spin-singlet pairing \cite{Zhao2022, Li2022}. \nh{The validity of the mean-field approximation in the band HK model was analyzed using the Ginzburg criterion in Ref.~\onlinecite{Zhao2022}. For the present two-dimensional two-orbital model, we perform a corresponding analysis in Appendix~\ref{A2} and find that the range of validity depends strongly on $U$ and $g$.} 

The mean-field Hamiltonian is
\begin{equation}
    \op{H} = \op{H}{\text{OHK}} - \frac{g}{N} \hc{A} \op{A}
    \approx \op{H}{\text{OHK}} - \Delta^* \op{A} - \Delta \hc{A} + \frac{N}{g} \abs{\Delta}^2,
\end{equation}
where the order parameter is given by the thermal average
\begin{equation}
    \Delta = \frac{g}{N} \left\langle \op{A} \right\rangle,
    \qquad
    \left\langle \cdot \right \rangle = \frac{1}{Z} \tr \cdot e^{-\beta H}.
\end{equation}
Since $\langle \op{A} \rangle$ is extensive in the system size, the order parameter $\Delta = g/N\, \langle \op{A} \rangle$ is intensive, i.e. $\Delta \sim \mathcal{O}(1)$.

The mean-field Hamiltonian couples opposite momenta $\mathbf{k}$ and $-\mathbf{k}$ and can thus be written as
\begin{equation}
    \op{H} = \frac{N}{g} \abs{\Delta}^2 + \sum_{\mathbf{k} \in \text{HBZ}} \op{H}{\mathbf{k}},
\end{equation}
where the sum runs over the half Brillouin zone containing one representative of each pair $\{\mathbf{k}, -\mathbf{k} \}$. The local \mbox{Hamiltonian}
\begin{equation}    \label{4LocHamiltonian}
\begin{split}
    \op{H}{\mathbf{k}} =& \sum_{s = \pm, \alpha, \beta, \sigma} \left( H_{\alpha \beta}(s \mathbf{k}) - \mu \delta_{\alpha \beta} \right) \hc{c}{s \mathbf{k} \alpha \sigma} \op{c}{s \mathbf{k} \beta \sigma} 
    \\
    &+ U \sum_{s = \pm, \alpha}  \op{n}{s \mathbf{k} \alpha \uparrow} \op{n}{s \mathbf{k} \alpha \downarrow} 
    \\
    &-\Delta^* \sum_{s = \pm} \op{\psi}{-s \mathbf{k}}{T} W(s \mathbf{k}) \op{\psi}{sk}
    - \Delta \sum_{s = \pm} \hc{\psi}{s \mathbf{k}} W^\dag(s \mathbf{k}) \hc{\psi}{-s \mathbf{k}}{T}
\end{split}
\end{equation}
acts on the many-body Fock space of the $4 n$ Fermionic modes $(\pm \mathbf{k}, \alpha, \sigma)$ and thus has dimension $2^{4n}$. For the two-band case ($n=2$), this yields a local Hilbert-space dimension of $2^8 = 256$. The partition function is
\begin{equation}
    Z = \tr\, e^{-\beta \op{H}} = e^{-\frac{\beta N}{g} \abs{\Delta}^2} \prod_{\mathbf{k} \in \text{HBZ}} Z_\mathbf{k},  \qquad  Z_\mathbf{k} = \tr_\mathbf{k} e^{-\beta \op{H}{\mathbf{k}}},
\end{equation}
where we denote the trace over a single $\mathbf{k}$-point as $\tr_\mathbf{k}$ and we call $Z_\mathbf{k}$ the local partition function. The free energy is
\begin{equation}
    F = -T \ln Z = \frac{N}{g} \abs{\Delta}^2 - T \sum_{\mathbf{k} \in \text{HBZ}} \ln\, Z_\mathbf{k}.
\end{equation}

\nh{
In the following, we consider the intensive free energy density $f = F/N$. For each pairing channel and for each set of parameters $U$, $g$, and $T$, we evaluate $f(\Delta)$ as a function of the order parameter $\Delta$, which is chosen real without loss of generality. We first determine the chemical potential from the normal-state problem, $\Delta = 0$, by imposing half filling. This value of $\mu$ is then kept fixed when evaluating $f(\Delta)$. Throughout this work, half filling refers to the normal state used to fix $\mu$. The density is not constrained to remain at half filling for $\Delta \neq 0$.

For each value of $\Delta$, the local Hamiltonian $\op{H}{\mathbf{k}}$ is diagonalized exactly in each momentum sector, and the corresponding local partition functions $Z_{\mathbf{k}}$ are used to construct $f(\Delta)$. The Brillouin-zone sums are evaluated on a symmetry-reduced shifted Monkhorst-Pack mesh \cite{MonkhorstPack1976} using 64 inequivalent sampling points and appropriate symmetry weights. Due to the shift, no time-reversal-invariant momenta occur, i.e. $\mathbf{k}$ and $-\mathbf{k}$ are distinct for all $\mathbf{k} \in \mathrm{HBZ}$. We checked representative parameter points in all four channels in Eq.~\eqref{3SelectedPairingChannels} using denser momentum-space meshes and found no qualitative change in the free-energy landscapes.

The extrema of $f(\Delta)$ are then tracked as a function of temperature. The physical solution is identified with the global minimum, while additional local extrema are included when discussing the free-energy landscape. The critical temperature $T_C$ is then defined as the temperature at which the global minimum changes between the normal-state solution $\Delta = 0$ and a superconducting solution $\Delta \neq 0$.
}

\nh{As a phenomenological choice of tight-binding parameters,} we choose the nearest-neighbour Slater-Koster integrals as $(pp \sigma)_1 = 1$ and $(pp \pi)_1 = -1/2$. The second-nearest neighbour hoppings are \nh{estimated} via a Harrison scaling with $1/a^\eta$, where $a = \sqrt{2}$ is the distance of second-nearest neighbours on the square lattice and $\eta = 2$ is a typical exponent for p-orbitals \cite{HarrisonBook}. This results in $(\rm{pp} \sigma)_2 = 1/2$ and $(\rm{pp} \pi)_2 = -1/4$. These values are kept fixed throughout all numerical calculations. The non-interacting bandwidth of our model is $W = 6$ with these parameters. We thus vary the interaction strength $U$ in $[0, 9]$.

We illustrate the structure of the mean-field solutions through representative free-energy landscapes. These examples provide the basis for identifying the critical temperature and highlight both conventional and anomalous behaviour in different pairing channels and for different values of the interaction strength.

\begin{figure}[t!]
  \centering
  \includegraphics[width=\linewidth]{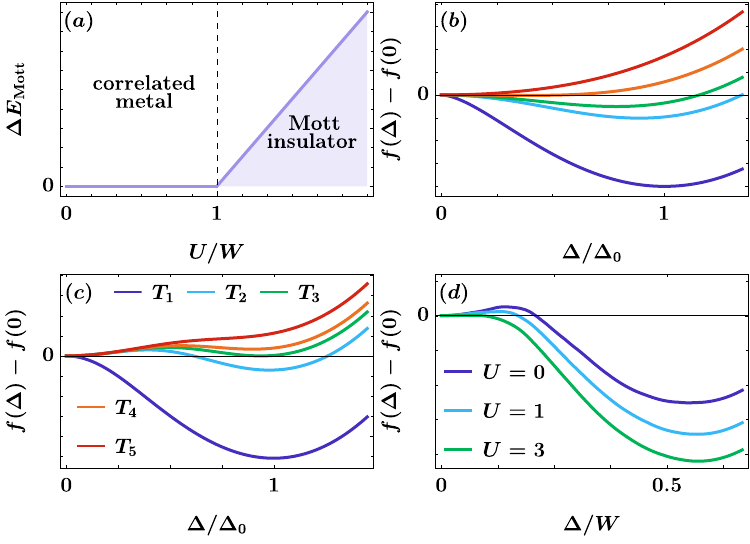}
  \caption{
  (a) Mott gap $\Delta E_\text{Mott}$ as a function of interaction strength $U/W$.
  (b) Difference in free energy density $f(\Delta) - f(0)$ for the pairing channel $\rm A_{1g}^{(ST)}$ at $U = 2$ and representative temperatures.
  (c) Difference in free energy density $f(\Delta) - f(0)$ for the same pairing channel at $U = 8$ and representative temperatures $T_1 < T_2 < T_3 < T_4 < T_5$.
  (d) Difference in free energy density $f(\Delta) - f(0)$ for the pairing channel $\rm E_u^{(SS)}$ at $T = 0$ and selected values of $U$.
  Panels (b)--(d) are shown for fixed pairing strength $g = 4$.
  In panels (b) and (c), $\Delta_0$ denotes the largest value of $\Delta$ shown in the corresponding panel.
  }
  \label{Fig_FreeEnergy}
\end{figure}

To provide a reference point for the interaction strength, we define the Mott gap from the zero-temperature normal-state density $\rho(\mu)$. In the Mott regime the density develops a plateau at half-filling. The width of this plateau defines the Mott gap
\begin{equation}    \label{4MottGap}
    \Delta E_\text{Mott} = \mu_+ - \mu_-, 
\end{equation}
where $\mu_\pm$ denote the upper and lower bounds of the Mott plateau, respectively. Figure~\ref{Fig_FreeEnergy} (a) shows the Mott gap as a function of the interaction strength $U$ in units of the non-interacting bandwidth $W = 6$ for the normal-state orbital HK model. The gap opens above a critical interaction strength $U_C = W$, which is the same critical value as in the band HK model.

\begin{figure}[t!]
  \centering
  \includegraphics[width=\linewidth]{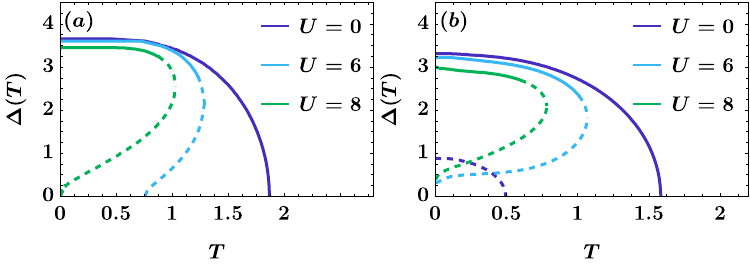}
  \caption{Positions of the extrema of the free energy as functions of temperature for the pairing channels $\rm A_{1g}^{(ST)}$ (a) and $\rm E_u^{(SS)}$ (b) for selected values of $U$ and at fixed pairing strength $g = 4$. Solid lines denote global minima, while dashed lines denote additional local extrema, including both local minima and local maxima. \nh{The extremum at $\Delta = 0$ is not shown.}}
  \label{Fig_OrderParameter}
\end{figure}

Figure~\ref{Fig_FreeEnergy} (b) and (c) display the free energy for the pairing channel $\rm A_{1g}^{(ST)}$ at $U = 2$, i.e. below the Mott transition, and at $U = 8$ within the Mott regime, respectively. \nh{For the example at $U = 2$,} the free energy exhibits the conventional BCS behaviour: below the critical temperature, the global minimum is located at $\Delta \neq 0$. Upon increasing temperature, this minimum moves continuously towards $\Delta = 0$ and eventually merges with the normal-state solution, corresponding to a conventional superconducting transition.

\nh{For the example at $U=8$, which lies in the Mott regime,} the global minimum remains at $\Delta \neq 0$ upon increasing the temperature and is separated from the normal-state solution at $\Delta = 0$ by a local maximum. Eventually, the system reaches a state with \nh{minima of equal free energy} at $\Delta = 0$ and $\Delta \neq 0$, marking the point of the first-order phase transition. Increasing the temperature further, the local minimum for $\Delta \neq 0$ persists as a metastable state even above the critical temperature. \nh{This behaviour} has also been described for the band HK model in Ref.~\onlinecite{Zhao2022}. We observe analogous behaviour for the pairing channels $\rm B_{2g}^{(TS)}$ and $\rm B_{1g}^{(ST)}$. 

\nh{The transition order across the $(U,g)$ plane is systematically analyzed in Appendix~\ref{A2}. The boundary between continuous and first-order transitions depends on both $U$ and $g$ and does not generally coincide with the normal-state Mott transition.}

\begin{figure}[t!]
  \centering
  \includegraphics[width=\linewidth]{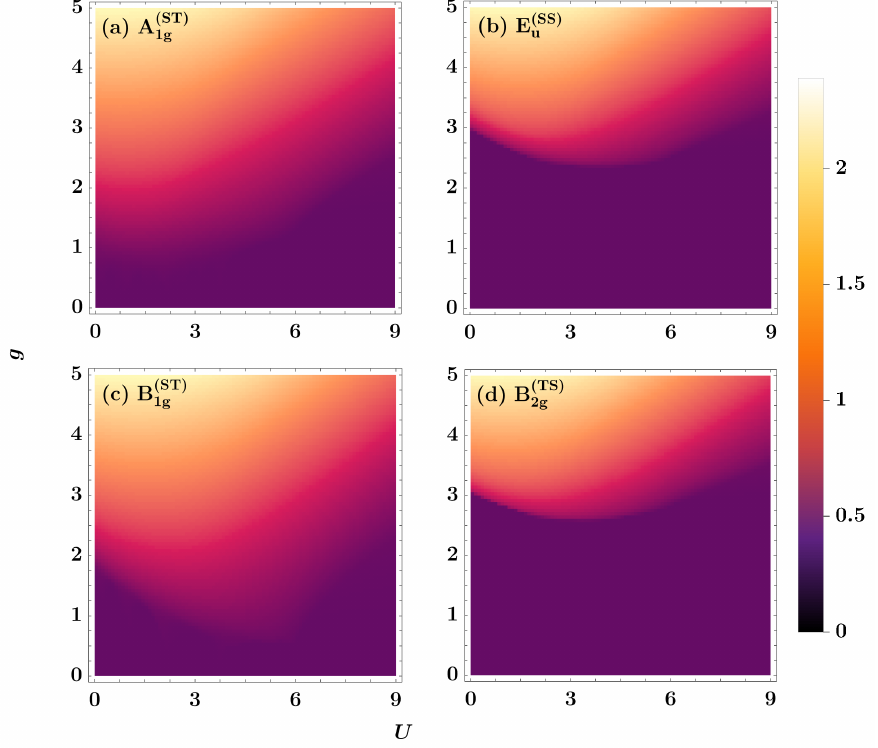}
  \caption{Critical temperature $T_C$ as a function of interaction strength $U$ and pairing strength $g$ for the four selected pairing channels given in Eq.~\eqref{3SelectedPairingChannels}: 
  (a) $\rm A_{1g}^{(ST)}$,
  (b) $\rm E_u^{(SS)}$,
  \nh{(c) $\rm B_{1g}^{(ST)}$ and 
  (d) $\rm B_{2g}^{(TS)}$.}
  }
  \label{Fig_TCHeatMap}
\end{figure}

\begin{figure}
    \centering
    \includegraphics[width=\linewidth]{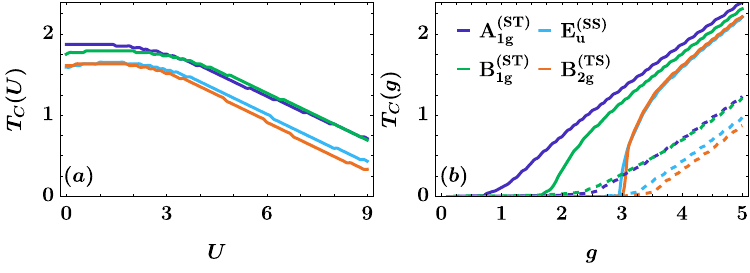}
    \caption{\nh{Critical temperature $T_C$ for the four selected pairing channels given in Eq.~\eqref{3SelectedPairingChannels}. (a) $T_C$ as a function of interaction strength $U$ at fixed pairing strength $g = 4$. (b) $T_C$ as a function of pairing strength $g$ at fixed $U = 0$ (solid lines) and $U = 9$ (dashed lines).
    }}
    \label{Fig_TCScaling}
\end{figure}

\begin{figure}
    \centering
    \includegraphics[width=\linewidth]{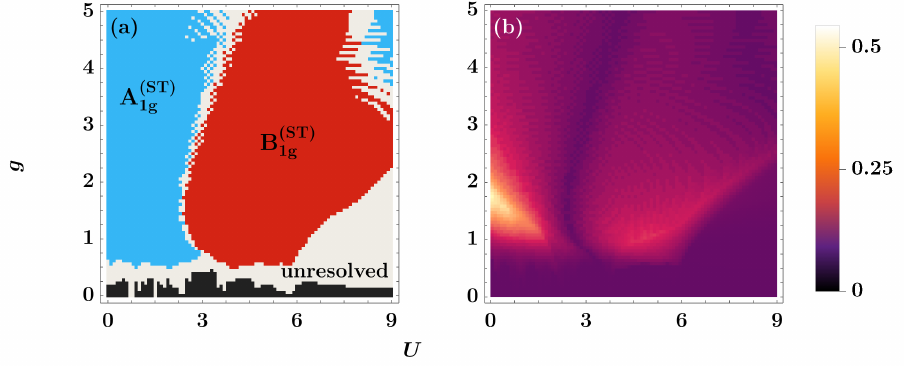}
    \caption{\nh{
    Comparison of the leading superconducting channel among the four selected gap functions in Eq.~\eqref{3SelectedPairingChannels}. (a) Leading-channel map in the $U$-$g$ plane. At each point the channel with the largest $T_C$ is shown. Grey regions indicate points where the two largest values of $T_C$ are numerically unresolved. Black regions indicate points where no finite $T_C$ is found among the four selected channels. (b) Difference $T_C^{(1)} - T_C^{(2)}$ between largest and second largest critical temperatures.
    }}
    \label{Fig_TCChannelComparison}
\end{figure}

Compared with the $\rm A_{1g}^{(ST)}$ example, the $\rm E_u^{(SS)}$ channel shown in Fig.~\ref{Fig_FreeEnergy} (d) displays a different free-energy topology. Here, a local maximum persists down to $T = 0$\nh{, which separates the superconducting minimum from the metastable normal-state minimum $\Delta = 0$. Its persistence at $U = 0$ shows that this barrier does not require the HK interaction. On the contrary,} the anomalous structure is gradually suppressed with increasing interaction strength and disappears in the Mott-regime parameter range shown here.

The different free-energy topologies are reflected in the extrema as functions of the temperature. Figure~\ref{Fig_OrderParameter} shows the positions of these extrema for the pairing channels $\rm A_{1g}^{(ST)}$ and $\rm E_u^{(SS)}$ at fixed pairing strength $g = 4$ and selected values of the interaction strength. For the $\rm A_{1g}^{(ST)}$ channel, the non-interacting case shows the expected continuous suppression of the minimum as the temperature is increased. \nh{For the larger interaction strengths shown, the order parameter drops discontinuously to zero at $T_C$, while additional local extrema persist over a finite temperature interval.} The $\rm E_u^{(SS)}$ channel shows a more intricate structure, including additional extrema even \nh{in the non-interacting limit.}

\nh{
In addition to these free-energy landscapes and the associated extrema, we show the full dependence of the critical temperature on interaction and pairing strength for the selected pairing channels in Fig.~\ref{Fig_TCHeatMap}. For the four pairing channels chosen in Eq.~\eqref{3SelectedPairingChannels}, the critical-temperature fields fall into two qualitative groups. The $\rm A_{1g}^{(ST)}$ and $\rm B_{1g}^{(ST)}$ channels in panels (a) and (c), both of which are spin-singlet channels, show a broad region of finite critical temperature already at comparatively small pairing strength. By contrast, the $\rm E_u^{(SS)}$ and $\rm B_{2g}^{(TS)}$ channels in panels (b) and (d) show strongly suppressed critical temperatures and reach larger values only for larger $g$. In all channels, increasing $g$ enhances the critical temperature, as expected. The critical temperature is maximized at finite interaction strength, with the maximum occurring well below the Mott transition.
}

\nh{
In Fig.~\ref{Fig_TCScaling}, we consider the dependence on $U$ and $g$ more explicitly. Panel (a) shows $T_C(U)$ for fixed $g = 4$. All four channels exhibit a similar qualitative dependence on $U$. The critical temperature is weakly enhanced at small interaction strength, reaches a maximum, and is subsequently suppressed with increasing $U$. Panel (b) shows $T_C(g)$ for fixed $U = 0$ and $U = 9$. The onset of finite $T_C$ depends strongly on the pairing channel and pairing strength. At larger pairing strength, however, $T_C$ increases approximately linearly with $g$ in all four channels, with similar but not identical slopes. Thus, while the channels share common qualitative trends, we do not find a single channel-independent scaling form for $T_C$.
}

\nh{
To compare the channels directly, we construct the leading-channel map shown in Fig.~\ref{Fig_TCChannelComparison}. In panel (a), we identify at each point in the $(U, g)$ plane the channel with the largest critical temperature among the four selected gap functions. For this comparison, we use the same value of the phenomenological pairing strength $g$ for all four channels. The comparison is restricted to the individual pairing channels defined in Eq.~\eqref{3SelectedPairingChannels}. Mixed or coexisting order parameters of different channels are not considered. Among these channels, the largest critical temperature is found in the $\rm A_{1g}^{(ST)}$ and $\rm B_{1g}^{(ST)}$ channels in the parameter range studied. The $\rm A_{1g}^{(ST)}$ channel is leading at smaller interaction strength, whereas the $\rm B_{1g}^{(ST)}$ channel becomes leading over a broad region at larger $U$. This corresponds to a change of the leading superconducting channel from the $\rm s$-wave $\rm A_{1g}^{(ST)}$ channel to the nodal $\mathrm{d}_{x^2-y^2}$-wave $\rm B_{1g}^{(ST)}$ channel. The $\rm E_u^{(SS)}$ and $\rm B_{2g}^{(TS)}$ channels remain subleading throughout this comparison. Grey regions indicate points where the two largest values of $T_C$ are numerically unresolved, while black regions indicate points where no finite $T_C$ is found among the four selected channels. Panel (b) shows the corresponding separation $T_C^{(1)} - T_C^{(2)}$ between the leading and subleading critical temperatures and thereby quantifies the robustness of the leading channel.
}

\section{Conclusion}
\label{secV}

We have studied superconductivity in a two-orbital extension of the HK model for p-orbitals on the square lattice. For this specific orbital setting based on the point group $\rm D_{4h}$, we provide a systematic classification of the symmetry-allowed superconducting basis functions in all spin-, orbital- and momentum sectors.

For selected pairing channels from this classification, we computed the mean-field free energy, obtained from exact diagonalization of the local momentum-sector Hamiltonians, and the critical temperature as functions of interaction strength and pairing strength. The HK interaction affects not only the critical temperature, but also the topology of the free energy. \nh{For the selected pairing channels, we find both continuous and first-order superconducting transitions depending on interaction and pairing strength. First-order transitions are accompanied by metastable states in the free-energy landscape. The boundary between continuous and first-order behavior does not generally coincide with the normal-state Mott transition.} In the channel $\rm E_u^{(SS)}$, a local maximum persists down to $T = 0$ in the metallic regime. The critical temperature shows a pronounced channel dependence and is maximized at finite interaction strength, with the maximum occurring well below the Mott transition. \nh{A direct comparison of the selected channels shows that the largest critical temperatures are found in the intraband $\rm A_{1g}^{(ST)}$ and $\rm B_{1g}^{(ST)}$ channels. This corresponds to an interaction-driven change of the leading superconducting channel from the fully symmetric, s-wave $\rm A_{1g}^{(ST)}$ channel to the nodal, $\mathrm{d}_{x^2-y^2}$-wave $\rm B_{1g}^{(ST)}$ channel.} \nh{In systems displaying analogous physics, the first-order transition would be reflected in a discontinuous change of the superconducting order parameter, while the accompanying metastable states could additionally give rise to hysteresis under temperature cycling \cite{RoyChaddah2004}.} 

While the classification derived here is specific to the p-orbital model on the square lattice, the HK construction itself is more general. \nh{The same classification procedure can be performed for other orbital representations, such as the $(\rm d_{xz}$, $\rm d_{yz})$-doublet, which transforms according to $\rm E_g$ instead of $\rm E_u$.} Momentum-local interactions can be implemented for other band structures and lattice geometries, including systems where the relevant degrees of freedom are not described by the same $\rm D_{4h}$ symmetry. \nh{One motivation for such extensions comes from two-dimensional superconducting metal-organic frameworks, where several metal- and ligand-derived orbitals may contribute near the Fermi energy. A quantitative application to a specific material would require the corresponding material-specific band structure and interaction terms.} \nh{Likewise, the present two-orbital calculation does not imply a general dependence of $T_C$ on the number of orbitals, since additional orbitals also modify the band structure and interaction and pairing channels.} \nh{Including momentum-mixing terms in the interaction would not invalidate the symmetry classification, provided that the point-group symmetry is preserved, but would require a different numerical treatment because the momentum-sector decomposition used here would no longer apply.} \nh{More general multiorbital extensions, including interorbital interactions described by off-diagonal elements of $U_{\alpha \beta}$ and additional momentum-local Hund-exchange terms, as well as pair-density waves, which have recently been discussed in the HK context \cite{Froldi2024}, are natural directions for future work.} The present work therefore provides a concrete example of how superconducting order can be classified and analyzed in an orbital HK model, while the broader framework can be adapted to other correlated multiband systems.

\begin{acknowledgments}
All authors acknowledge support from the Knut and Alice Wallenberg Foundation (Grant No. 2023.0087), the Swedish Research Council (VR starting Grant No. 2022-03350), the Olle Engkvist Foundation (Grant No. 229-0443), and Chalmers University of Technology, via the department of physics and the Areas of Advance Nano and Materials Science. The computations were enabled by resources provided by the National Academic Infrastructure for Supercomputing in Sweden (NAISS) at C3SE partially funded by the Swedish Research Council through grant agreement no. 2022-06725.
\end{acknowledgments}

\bibliography{lit}
\appendix

\section{Two-Band Model}
\label{A1}

\begin{figure}[!h]
    \centering
    \includegraphics[width=\linewidth]{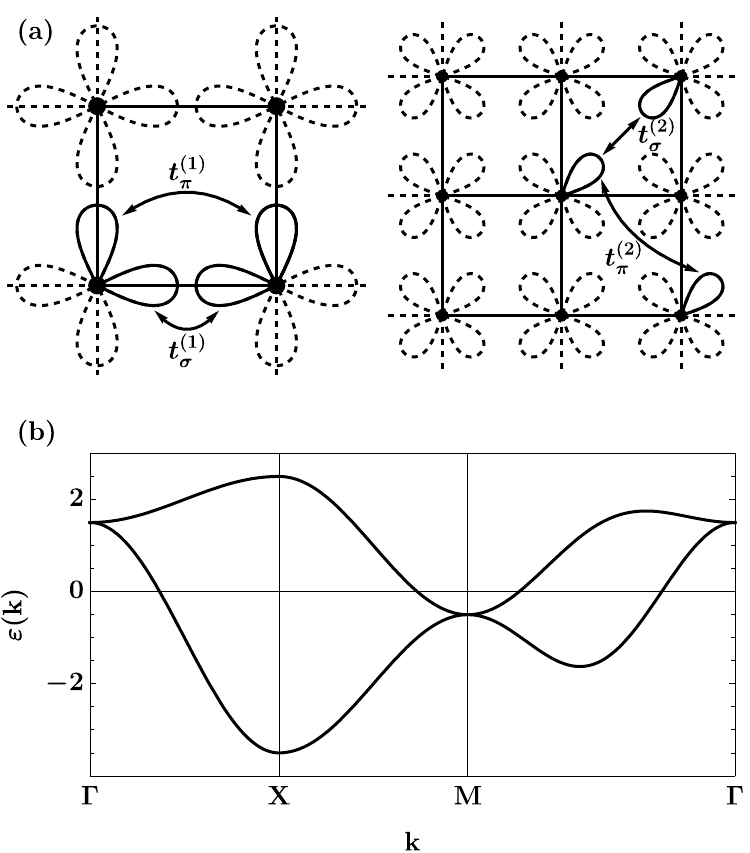}
    \caption{(a) Illustration of nearest- and second-nearest-neighbour hopping between p-orbitals on the square lattice. (b) Band structure of the two-band model along an irreducible path in the square Brillouin zone.}
    \label{A1BandStructure}
\end{figure}

The two-band model used in the main text is a tight-binding model for $\mathrm{p}_x$- and $\mathrm{p}_y$-orbitals on the square lattice. It includes nearest- and second-nearest-neighbour hopping, and respects the point-group symmetry $\rm D_{4h}$. For notational simplicity, we suppress the spin index throughout this appendix. The tight-binding Hamiltonian is spin-degenerate and thus acts trivially in spin space.

The real space Hamiltonian is
\begin{equation}
\begin{split}
    \op{H} &= \sum_{i,j} \left[ t_\sigma^{(1)} \left( \hc{c}{\mathrm{p}_x i j} \op{c}{\mathrm{p}_x i+1 j} + \hc{c}{\mathrm{p}_y i j} \op{c}{\mathrm{p}_y i j+1} \right) \right.
    \\
    &+ \left. t_\pi^{(1)} \left( \hc{c}{\mathrm{p}_x i j} \op{c}{\mathrm{p}_x i j+1} + \hc{c}{\mathrm{p}_y i j} \op{c}{\mathrm{p}_y i+1 j} \right) + \text{h.c.} \right]
    \\
    &+ \sum_{i,j} \left[ t_\sigma^{(2)} \left( \hc{c}{\mathrm{p}_+ i j} \op{c}{\mathrm{p}_+ i+1 j+1} + \hc{c}{\mathrm{p}_- i j} \op{c}{\mathrm{p}_- i+1 j-1} \right) \right.
    \\
    &+ \left. t_\pi^{(2)} \left( \hc{c}{\mathrm{p}_+ i j} \op{c}{\mathrm{p}_+ i+1 j-1} + \hc{c}{\mathrm{p}_- i j} \op{c}{\mathrm{p}_- i+1 j+1} \right) + \text{h.c.} \right],
\end{split}
\end{equation}
where $\op{c}{\mathrm{p}_xij}$ and $\op{c}{\mathrm{p}_yij}$ denote the electronic annihilation operator for the $\mathrm{p}_x$- and $\mathrm{p}_y$- orbitals at site $i, j$, respectively. The nearest-neighbour terms distinguish between $\sigma$- hopping along the direction of the orbital and $\pi$- hopping perpendicular to it. For the second nearest-neighbour hopping, it is convenient to introduce hybridized orbitals oriented along the diagonal direction
\begin{equation}
    \op{c}{\mathrm{p}_\pm ij} = \frac{\op{c}{\mathrm{p}_x ij} \pm \op{c}{\mathrm{p}_y ij}}{\sqrt{2}}.
\end{equation}
The hopping processes are illustrated in Fig.~\ref{A1BandStructure}(a).

The model parameters are the Slater-Koster overlap integrals for $\sigma$- and $\pi$-bonds \cite{SlaterKoster1954}
\begin{alignat}{2}
    t_\sigma^{(1)} = (\rm{pp}\sigma)_1,   \qquad
    t_\pi^{(1)} = (\rm{pp}\pi)_1,
    \\  \nonumber
    t_\sigma^{(2)} = (\rm{pp}\sigma)_2 ,  \qquad
    t_\pi^{(2)} = (\rm{pp}\pi)_2.
\end{alignat}
The subscript denotes the neighbour shell and $\sigma$ and $\pi$ specify the orbital orientation relative to the hopping direction.

In momentum space we obtain
\begin{equation}
\begin{split}
    \op{H} =& \sum_\mathbf{k} \left( \hc{c}{\mathrm{p}_x \mathbf{k}}\, \hc{c}{\mathrm{p}_y \mathbf{k}} \right) H(\mathbf{k}) 
    \begin{pmatrix}
    \op{c}{\mathrm{p}_x \mathbf{k}} \\ \op{c}{\mathrm{p}_y \mathbf{k}}
    \end{pmatrix},
    \\
    H(\mathbf{k}) =& \left[ 4s \cos k_x \cos k_y + \left( t_\sigma^{(1)} + t_\pi^{(1)} \right) \left( \cos k_x + \cos k_y \right) \right] \tau_0 
    \\
    &- 4r \sin k_x \sin k_y \tau_x + \left( t_\sigma^{(1)} - t_\pi^{(1)} \right) \left( \cos k_x - \cos k_y \right) \tau_z,
\end{split}
\end{equation}
where $\tau_x$, $\tau_y$, and $\tau_z$ are the Pauli matrices in orbital space and $\tau_0$ is the identity matrix. We introduced combinations
\begin{equation}
    s = \frac{(\rm{pp}\sigma)_2 + (\rm{pp}\pi)_2}{2},  
    \qquad
    r = \frac{(\rm{pp}\sigma)_2 - (\rm{pp}\pi)_2}{2}.
\end{equation}
The parameter $s$ enters the orbital-independent part proportional to $\tau_0$, whereas $r$ enters the off-diagonal part and therefore mixes the $\mathrm{p}_x$- and $\mathrm{p}_y$-components.

Diagonalizing $H(\mathbf{k})$ gives the two bands
\begin{equation}
\begin{split}
    &\varepsilon_{1,2}(\mathbf{k}) = \left( t_\sigma^{(1)} + t_\pi^{(1)} \right) (\cos k_x + \cos k_y ) + 4s \cos k_x \cos k_y
    \\
    &\pm \sqrt{16\, r^2 \sin^2 k_x \sin^2 k_y + \left( t_\pi^{(1)} - t_\sigma^{(1)} \right)^2 (\cos k_x - \cos k_y)^2},
\end{split}
\end{equation}
which become degenerate at the $\Gamma$- and $\rm M$-point, where the orbital splitting terms vanish by symmetry. The resulting band structure along an irreducible path in the square Brillouin zone is shown in Fig.~\ref{A1BandStructure}.

\nh{\section{Ginzburg Criterion and Validity of Mean-Field Theory}
\label{A2}

\begin{figure}[t!]
  \centering
  \includegraphics[width=\linewidth]{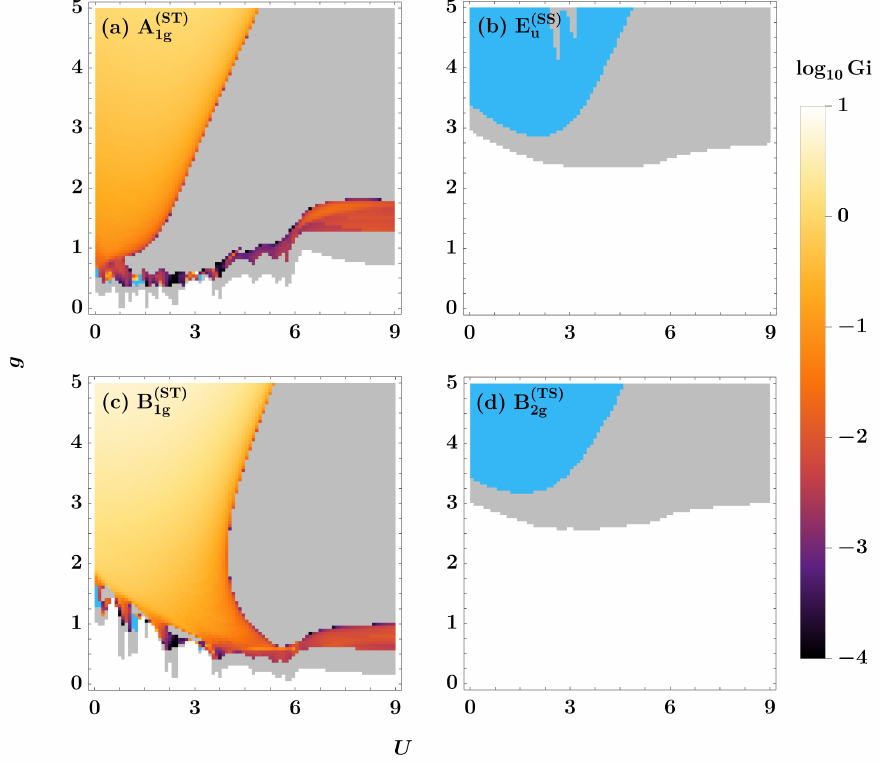}
  \caption{\nh{Ginzburg number as a function of interaction strength $U$ and pairing strength $g$ for the four selected pairing channels given in Eq.~\eqref{3SelectedPairingChannels}. The color plot shows $\log_{10}\mathrm{Gi}_{2\mathrm{D}}$ for continuous transitions with a zero-momentum critical mode. Gray regions indicate first-order transitions, blue regions indicate negative gradient coefficients along at least one crystallographic direction at the critical temperature, and white regions indicate no homogeneous superconducting transition.}}
  \label{A2Ginzburg}
\end{figure}

To explore the validity of the mean-field approximation, we evaluate the Ginzburg criterion for the continuous superconducting transitions. We expand the free energy in the order parameter and its spatial derivatives within Ginzburg-Landau theory. The required coefficients are extracted numerically from the free-energy landscape of the two-orbital HK model.

We introduce the pairing operator and the order parameter for Cooper pairs with finite center-of-mass momentum $\mathbf{q}$
\begin{equation}
    \op{A}{\mathbf{q}} = \sum_\mathbf{k} \op{\psi}{-\mathbf{k} + \mathbf{q}/2}{T} W(\mathbf{k}) \op{\psi}{\mathbf{k} + \mathbf{q}/2},
    \qquad
    \Delta_\mathbf{q} = \frac{g}{N} \left\langle \op{A}{\mathbf{q}} \right\rangle.
\end{equation}
The Hamiltonian for a single momentum $\mathbf{q}$ is
\begin{equation}
    \op{H}(\mathbf{q}) = \op{H}{\text{OHK}} - \frac{g}{N} \hc{A}{\mathbf{q}} \op{A}{\mathbf{q}} 
    \approx \op{H}{\text{OHK}} - \Delta_\mathbf{q}^* \op{A}{\mathbf{q}} - \Delta_\mathbf{q} \hc{A}{\mathbf{q}} + \frac{N}{g} \abs{\Delta_\mathbf{q}}^2,
\end{equation}
where we assume that the pairing strength $g$ is $\mathbf{q}$-independent and perform the mean-field decoupling as in Sec.~\ref{secIV}.

The Ginzburg-Landau free-energy functional is given by
\begin{equation}
\begin{split}
    F_\text{GL}[\Delta] = F_\text{N} 
    + \int d^2r \left[
    a(T) \abs{\Delta(\mathbf{r})}^2 
    + \frac{b(T)}{2} \abs{\Delta(\mathbf{r})}^4
    \right.
    \\
    \left.
    + \sum_{i=x,y} K_i(T) \abs{\partial_i \Delta(\mathbf{r})}^2 + \ldots \right],
\end{split}
\end{equation}
where $x$ and $y$ are oriented along the crystallographic axes. For the one-dimensional irreducible representations, the order parameter is scalar, and $\rm D_{4h}$ symmetry implies $K_x(T) = K_y(T)$, while mixed gradient terms vanish. For the two-dimensional representation $\rm E_u$, the order parameter has two components and allows a more general gradient structure. Since we restrict the analysis to a single $p_x$-like component, we retain the scalar form, but allow for distinct stiffness coefficients $K_x(T)$ and $K_y(T)$.

The real- and momentum-space representations of the order parameter are related by the Fourier transform
\begin{equation}
    \Delta(\mathbf{r}) = \sum_\mathbf{q} \Delta_\mathbf{q}\, e^{i\mathbf{q} \cdot \mathbf{r}},
    \qquad
    \Delta_\mathbf{q} = \frac{1}{N} \int d^2r\, \Delta(\mathbf{r}) e^{-i\mathbf{q} \cdot \mathbf{r}},
\end{equation}
where the lattice constant, and hence the unit-cell area, is set to unity. As in the main text, we denote the free energy density as $f_\text{GL} = F_\text{GL}/N$. Switching to momentum space allows us to define a $\mathbf{q}$-dependent quadratic coefficient
\begin{equation}
\begin{split}
    f_\text{GL}^{(2)}[\Delta] &= \sum_\mathbf{q} a(\mathbf{q}, T) \abs{\Delta_\mathbf{q}}^2,
    \\
    a(\mathbf{q}, T) &= a(T) + \sum_{i=x,y} K_i(T)\, q_i^2 + \mathcal{O} \left( \abs{\mathbf{q}}^4 \right).
\end{split}
\end{equation}
It describes the quadratic free-energy cost of a pairing mode with finite center-of-mass momentum $\mathbf{q}$. The homogeneous coefficient is recovered by $a(T) = a(\mathbf{0}, T)$.

The Ginzburg-Landau coefficients can be obtained by expanding the free energy around the normal state $\Delta = 0$
\begin{equation}
\begin{split}
    f(\Delta, \mathbf{q}, T) &= f(0, T) + a(\mathbf{q}, T) \abs{\Delta}^2 
    + \mathcal{O}\left( \abs{\Delta}^4 \right),
    \\
    f(\Delta, \mathbf{0}, T) &= f(0, T) + a(T) \abs{\Delta}^2 
    + \frac{b(T)}{2} \abs{\Delta}^4
    + \mathcal{O}\left( \abs{\Delta}^6 \right).
\end{split}
\end{equation}
As in the main text, we choose the scalar order-parameter amplitude $\Delta$ to be real. The coefficients are therefore obtained as
\begin{equation}
\begin{split}
    a(T) &= \frac{1}{2} \frac{\partial^2}{\partial \Delta^2} f(\Delta, \mathbf{0}, T) \Bigg|_{\Delta = 0},
    \\
    b(T) &= \frac{1}{12} \frac{\partial^4}{\partial \Delta^4} f(\Delta, \mathbf{0}, T) \Bigg|_{\Delta = 0},
    \\
    K_i(T) &= \lim_{q \to 0} \frac{a(q\, \hat{\mathbf{e}}_i, T) - a(\mathbf{0}, T)}{q^2}.
\end{split}
\end{equation}

For a continuous transition, the quadratic coefficient vanishes at the mean-field critical temperature, $a(T_C) = 0$, and the quartic coefficient remains positive, $b(T_C) > 0$. Close to $T_C$, the quadratic coefficient can be linearized
\begin{equation}
\begin{split}
    a(T) &= a'(T_C) (T - T_C) + \mathcal{O} \left( (T - T_C)^2 \right),
    \\
    a'(T_C) &= \frac{\partial}{\partial T} a(T) \Bigg|_{T = T_C}.
\end{split}
\end{equation}
To distinguish continuous and first-order transitions within the Landau analysis, we define the temperature $T_0$ by $a(T_0) = 0$. For $b(T_0) > 0$, the transition is continuous and $T_0 = T_C$. For $b(T_0) < 0$, the transition is of first order and, in general, $T_0 \neq T_C$, while $b(T_0) = 0$ marks the boundary between the two regimes.

The Ginzburg criterion estimates the region around the mean-field critical temperature in which thermal fluctuations become important. The two-dimensional Ginzburg number is given by \cite{Shanenko2022, LarkinVarlamovBook}
\begin{equation}
    \mathrm{Gi}_{2\mathrm{D}} = \frac{b(T_C)}{4\pi\, a'(T_C)\, \sqrt{K_x(T_C) K_y(T_C)}}.
\end{equation}
We use units with $k_\mathrm{B} = 1$. Mean-field theory is expected to break down when
\begin{equation}
    \frac{\abs{T - T_C}}{T_C} \lesssim \mathrm{Gi}_{2\mathrm{D}}.
\end{equation}
A small Ginzburg number therefore corresponds to a narrow fluctuation regime and a broad range of validity of the mean-field approximation.

For the numerical evaluation, we determine the coefficients entering the Ginzburg number at each point in the $(U, g)$ plane for which the superconducting transition is continuous. The quartic coefficient $b(T_C)$ is obtained from the homogeneous free-energy landscape, while $a'(T_C)$ follows from the temperature dependence of the quadratic coefficient in the vicinity of $T_C$. The stiffness coefficients $K_i(T_C)$ are extracted from the dependence of $a(\mathbf{q}, T_C)$ on small center-of-mass momenta along the crystallographic directions.

The conventional Ginzburg criterion applies only to continuous transitions for which the critical mode occurs at zero center-of-mass momentum. We therefore do not assign a Ginzburg number to first-order transitions. Likewise, if $K_i(T_C) < 0$ along at least one crystallographic direction, the quadratic coefficient $a(\mathbf{q}, T_C)$ decreases away from $\mathbf{q} = 0$, signaling an instability towards pairing at finite center-of-mass momentum. In this case, the homogeneous $\mathbf{q} = 0$ expansion does not describe the critical mode and the Ginzburg number defined above is not evaluated.

Figure~\ref{A2Ginzburg} shows the resulting Ginzburg numbers for the four pairing channels considered in the main text. For the $\rm A_{1g}^{(ST)}$ and $\rm B_{1g}^{(ST)}$ channels, the Ginzburg number varies by several orders of magnitude across the continuous-transition regime. In parts of the phase diagram close to
the lower boundary of the superconducting region, we find $\mathrm{Gi}_{2\mathrm{D}} \ll 1$, indicating that the temperature interval around $T_C$ where fluctuations dominate is narrow and that the mean-field description is well controlled away from this interval. With increasing pairing strength, however, the Ginzburg number generally increases and reaches values of order unity and larger. In these regimes, fluctuations are no longer small and the mean-field description of the critical
region is therefore not controlled.

For the $\rm E_u^{(SS)}$ and $\rm B_{2g}^{(TS)}$ channels, extended regions instead exhibit an instability of the homogeneous $\mathbf{q} = 0$ state towards pairing at finite center-of-mass momentum. These regions lie outside the homogeneous pairing ansatz considered in
the main text and would require a separate analysis allowing for finite-momentum superconducting order.

}

\raggedbottom

\end{document}